\documentclass{aa}
\usepackage{graphicx}
\usepackage{txfonts}
\usepackage{natbib}
\usepackage{amssymb}
\usepackage{amsfonts}
\bibpunct{(}{)}{;}{a}{}{,}

\newcommand{\cm}{cm$^{-2}$}
\newcommand{\kms}{km~s$^{-1}$}
\newcommand{\zem}{$z_{\rm em}$}

\newcommand{\HI}{\mbox{H\,{\sc i}}}
\newcommand{\Lya}{Lyman-$\alpha$}

\newcommand{\MgII}{\mbox{Mg\,{\sc ii}}}
\newcommand{\MgI}{\mbox{Mg\,{\sc i}}}
\newcommand{\SiI}{\mbox{Si\,{\sc i}}}
\newcommand{\CaI}{\mbox{Ca\,{\sc i}}}
\newcommand{\CaII}{\mbox{Ca\,{\sc ii}}}
\newcommand{\MnII}{\mbox{Mn\,{\sc ii}}}
\newcommand{\FeI}{\mbox{Fe\,{\sc i}}}
\newcommand{\FeII}{\mbox{Fe\,{\sc ii}}}

\begin{document}
  \title{A cold metal-poor cloud traced by a weak \MgII\ absorption at
        $z\simeq0.45$\thanks{Based on observations collected at the  
  European Southern Observatory Very Large Telescope, Cerro Paranal,
  Chile -- Program 166.A-0106(A) }}
   \subtitle{First detection of \SiI, \CaI, and \FeI\ in a QSO
        absorber} 
 \titlerunning{A cold metal-poor cloud at $z\simeq0.45$}

   \author{Valentina D'Odorico\inst{}}
           
   \offprints{V. D'Odorico}

   \institute{Osservatorio Astronomico di Trieste, via G.B. Tiepolo 11, I-34143 Trieste\\
              \email{dodorico@oats.inaf.it}}

\date{}

\abstract
{}
 { We present the observations of a weak \MgII\ absorption system detected 
   at $z\sim 0.452$ in the UVES high-resolution spectrum of the 
   QSO HE0001-2340. 
  The weaker of the two \MgII\ components forming the system 
  shows associated absorptions due to \SiI, \CaI, and  \FeI\ observed for the 
  first time in a QSO spectrum. 
  We investigate the nature of this absorber by comparing its properties 
  with those of different classes of absorbers and reproducing its
   ionization conditions with photoionization models. }
  {We measured column densities of \MgI, \MgII, \SiI, \CaI, \CaII,
  \MnII, \FeI, and \FeII\ with Voigt profile fitting. Although most of
  the observed velocity profiles  are not resolved
  in the UVES spectrum, a curve of growth analysis excluded a significant
  underestimation of the column densities, in particular for the \FeII\
  and \MgII\ multiplets.  
  We compared our measurements with a sample of 28 weak \MgII\
  systems detected in the interval $0.4 <z<1.4$ in the 18 UVES spectra of the ESO 
  Large Programme, 
  plus 24 weak \MgII\ systems 
  in the same redshift range taken from the literature. Then, we
  performed a comparison with 11 damped systems for
  which ionic column densities were measured component by component
  and with 44 Galactic lines of sight with measured \CaI, \CaII\ and \FeI. 
  We also ran a grid of photoionization models with
  Cloudy to reproduce the observed \MgI/\MgII, \CaI/\CaII, and
  \FeI/\FeII\ column density ratios for the studied system.}
 { The observed absorber belongs to the class of weak \MgII\
  systems on the basis of its equivalent width, however the relative
  strength of commonly observed transitions deviates significantly
  from those of the above mentioned absorbers. A rough estimate of the
  probability of crossing such a system with a QSO line of sight is
  $P\sim 0.03$.  
  The presence of rare neutral transitions suggests that the cloud is
  shielded by a large amount of neutral hydrogen.
  A detailed comparison of the observed column densities with the
  average properties of damped \Lya\ systems and local interstellar
  cold clouds shows, in particular, deficient \MgII/\MgI\ and
  \CaII/\CaI\ ratios in our cloud. 
  The results of photoionization models indicate that the cloud could be
  ionized by the UV background. However, a simple model of a single
  cloud with uniform density cannot reproduce the observed ionic
  abundance ratios, suggesting a more complex density structure for
  the absorber.
  Supposing that ionization corrections are negligible, the
  most puzzling result is the underabundance of magnesium with respect
  to iron which is hard to explain both with nucleosynthesis and with
  differential dust depletion. 
   }
{}

   \keywords{intergalactic medium, quasars: absorption lines}

   \maketitle


\section{Introduction}

Metal absorption systems observed in the spectra of high-redshift QSOs
are extremely useful for studying the nature and evolution of star
formation and gas chemical pollution through the cosmic ages.  

Damped \Lya\ systems (DLA) are the highest neutral-hydrogen column
density absorbers observed in QSO spectra ($N($\HI$)\ge 2 \times
10^{20}$ \cm).  
Such an amount of neutral gas is usually measured in local spiral
discs, and it ensures an extremely precise determination of the
chemical abundances due to the negligible ionization corrections
\citep{vladilo}.   
Cold neutral clouds could also be the precursors of molecular clouds,
the birthplace of stars. Indeed, molecular hydrogen has been observed
in 13-20 percent of the DLA \citep{srianand05}, and    
DLA at $z\sim 3.0-4.5$ contain enough mass in neutral gas to
account for a significant fraction of the visible stellar mass in
modern galaxies \citep[e.g.][]{storrielwolfe}. 

DLA are selected mainly in the optical by looking for absorption
features with equivalent width (EW) exceeding 5 \AA\ in normalised QSO
spectra \citep[][ and references therein]{prochaska05}.   
At redshifts $z\la1.6$, the \Lya\ line falls in the UV, requiring QSO
observations from space; furthermore, at low z  the interception
probability per unit redshift is reduced. An efficient technique for 
collecting large samples of low-redshift DLA is to pre-select lines of
sight showing \MgII\ absorptions and then do the follow-up
spectroscopy in the UV \citep{rao95}.  
The most recent observations by \citet{rao06} show that the fraction
of DLA in a  \MgII\ sample increases with the \MgII\ rest EW above a
threshold value of $W_o^{\lambda 2796}=0.6$ \AA. On the 
other hand, there is no correlation between $W_o^{\lambda 2796}$ and
$N($\HI$)$.  
The reasonable explanation given by Rao and collaborators is that the
largest EW systems arise in clouds bounded in
galaxy-sized potentials. A DLA system is observed if at least one of
the clouds along the line of sight is cold (less than 100 K) and has
a small velocity dispersion (a few \kms), so the larger the number of
clouds along the sightline, the higher the probability of encountering
a DLA system. Only rarely would a sightline intersect a single cloud
resulting in small $W_o^{\lambda 2796}$ and large $N($\HI$)$.  

In this work, we present the discovery of a peculiar absorption
system identified by a weak \MgII\ doublet at $z\simeq 0.452$ along the
line of sight to the bright QSO \object{HE0001-2340}.  
We compare its properties with those of weak \MgII, DLA systems 
and with local diffuse clouds. We claim that we have detected a first 
example of a cold intergalactic cloud and discuss the implication of this 
finding.  
The structure of the paper is the following: Sect.~2 briefly presents
the observational data and the reduction process; in Sect.~3 we report
the results of the fit of the lines and the comparison of the obtained
column densities and EWs with 3 samples of absorbers, 
each one presenting some characteristic in common with our system. In
Sect.~4 we report the results from a grid of photoionization models run
with Cloudy and we infer a physical model for the cloud. Finally,
Sect.~5 is dedicated to the concluding remarks.  
  				   

\section{Observations and data reduction}

The QSO HE0001-2340 (\zem~$=2.267$) was observed with UVES
\citep{dekker00} at the ESO VLT in the context of the ESO Large
Programme `The cosmic evolution of the IGM' \citep{bergeron04} at a
resolution of $R\simeq 45000$ and signal-to-noise ratio $S/N \simeq
60-100$ per pixel.  
The wavelength range goes from 3100 to 10000 \AA, except for the
intervals $\sim 5760-5830$ and $8515-8660$ \AA\ where the signal is
absent due to the gap between the two CCDs forming the red mosaic. 

The observations were reduced with the UVES pipeline \citep {ball00}
in the context of 
the ESO MIDAS data reduction package, applying the optimal extraction
methods and following the pipeline reduction step by step. Wavelengths
were corrected to vacuum-heliocentric values, and individual 1D spectra
were combined using a sliding window and weighting the signal by the
total errors in each pixel.  

The continuum level was determined by manually selecting regions of
the spectrum free of evident absorptions, which were  successively
fitted with a 3rd-degree spline polynomial.  
Finally the spectrum was divided by this continuum, leaving only the
information relative to absorption features.


\section{Analysis}

   \begin{figure}
   \centering
      \includegraphics[width=9truecm]{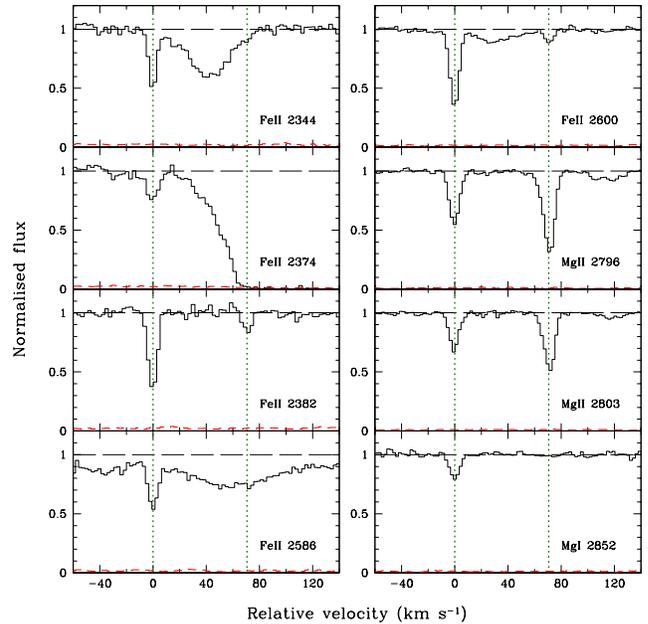}
     \caption{Transitions of \FeII, \MgII, and \MgI\ for the two
     velocity components of the discussed system. Velocity in the x
     axes is centered at the component which is discussed in the text
     ($z\simeq 0.4521$).}
         \label{fig:fe2mg2}
   \end{figure}

   \begin{figure}
   \centering
      \includegraphics[width=9truecm]{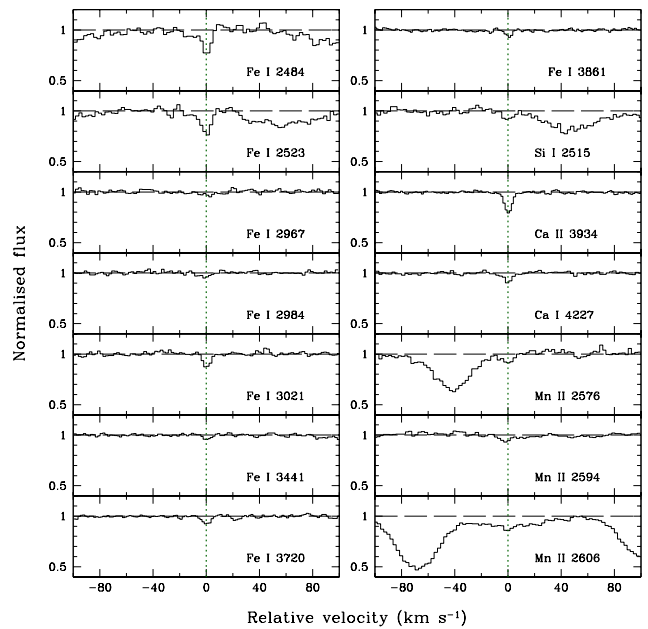}
     \caption{Other ionic transitions associated with the peculiar
     velocity component at $z\simeq 0.4521$. \FeI, \SiI\ and \CaI\ were
     observed for the first time in a QSO absorption system.}
         \label{fig:sys045}
   \end{figure}


In the process of detecting the metal lines contaminating the
Lyman-$\alpha$ forest, we have identified an \MgII\ doublet at
$z\sim0.452$ with the associated \FeII\ absorption, composed of two
extremely narrow components (see Fig.~\ref{fig:fe2mg2}).     

Surprisingly, at the redshift of the weaker \MgII\ component, $z
\simeq 0.45206$, we identified some ionic transitions generally
observed in stronger \MgII\ and DLA systems: \MgI\ $\lambda\, 2852$,
the triplet of \MnII\ at $\sim 2600$ and \CaII\ $\lambda\, 3934$  \AA\
(the other line of the doublet and the \mbox{Na\,{\sc i}} doublet fall
in the gap of the red CCDs).  
Even more exceptional is the detection of \CaI\ $\lambda\, 4227$,
\SiI\ $\lambda\, 2515$ \AA, and several transitions due to
\FeI. Those absorption lines are also weak in the local interstellar
medium, and this is the first time that they are detected associated
with a QSO absorber. 
All the detected absorption lines are shown in Figs.\ \ref{fig:fe2mg2}
and \ref{fig:sys045}. 

Voigt profiles were fitted to the observed absorption lines using the
context Lyman \citep{font:ball} of the ESO MIDAS package. 
The velocity profiles of the stronger
transitions, \MgII\ and \FeII\ $\lambda\,2382$, 2600 \AA, do not show
any sign of blending, and their Doppler parameters are comparable to
those found from the analysis of \MgII\ and \FeII\ absorption systems
in very high resolution QSO spectra \citep[$R\sim  120000$,][]{narayanan,chand06}.
The coincidence in redshift between different ions is of the order of 1
\kms\ suggesting that neutral and singly ionized elements occupy the
same region.  The resulting Doppler parameters are all below $\sim 4$
\kms, even below 1 \kms\ in the case of \FeI, implying that those
lines are not resolved at the resolution of our spectrum (FWHM~$\sim
6.7$ \kms).  
As a consequence, we could underestimate the column densities of the
stronger transitions, if they are saturated.     

   \begin{figure}
      \includegraphics[width=9truecm]{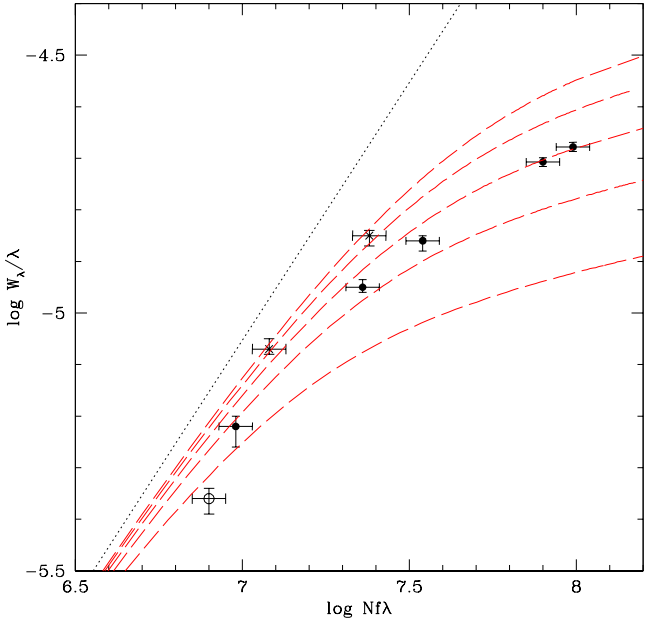}
     \caption{Data points obtained from the Voigt fit of \FeII\
     observed transitions (solid dots), the \MgII\ doublet (crosses)
     and the line \CaII\ $\lambda\,3934$ \AA\ (open dot) compared with
     the theoretical curve of growths computed for $b=3, 2.5, 2, 1.5, 1$
     \kms\ 
     (dashed lines from top to bottom). The dotted line represent the
     linear region of the curve of growth}
         \label{fig:cog}
   \end{figure}

In order to cross-check the results of our Voigt fit, a curve of
growth analysis has been performed with the multiplet of \FeII\ and
the \MgII\ doublet using EWs and column densities
determined with MIDAS and comparing them with theoretical curve of
growths at different Doppler parameters. 

It is clear from Fig.~\ref{fig:cog} that the computed column densities
are not significantly underestimated. On the contrary, they agree
satisfactorily with the curve of growths with Doppler parameters
between $\sim 1.5$ and 3 \kms.  We also plot the result for \CaII\
$\lambda\,3934$ \AA, which is, however, more uncertain since we do not  
observe the other line of the doublet.


%
\begin{table}
\caption{Fit of the observed transitions and limits on
  the rest EWs for some transitions commonly observed in
  the cold interstellar medium.}             
\label{linepar}      
\centering          
\begin{tabular}{c c c c }     
\hline\hline
&&&       \\
Transition & $W_{\lambda}$ & $\log N$ & $b$ \\ 
           & m\AA          & cm$^{-2}$ & \kms \\
\hline
\MgII\ $\lambda\,2796$ & $39\pm2$  & $12.15\pm0.03$ &$2.6\pm0.1$ \\
\MgII\ $\lambda\,2803$ & $24\pm2$  & $12.15\pm0.03$ & $2.6\pm0.1$\\
\MgI\ \ $\lambda\,2852$ & $18\pm2$ & $11.19\pm0.02$ & $2.9\pm0.4$ \\
\SiI\ \ \ $\lambda\,2515$ & $8\pm3$  & $11.84\pm0.05$ & $3.7\pm0.9$ \\
\CaII\ \ $\lambda\,3934$ & $17\pm1$ & $11.51\pm0.05$ & $1.0\pm 0.4$ \\
\CaI \ \ \ $\lambda\,4227$ & $11\pm3$ & $10.60\pm0.05$ & $1.5\pm0.5$ \\
\MnII\ $\lambda\,2576$ & $7\pm3$ & $11.56\pm0.06$ &  $3.7\pm0.9$ \\
\MnII\ $\lambda\,2594$ & $6\pm3$ & $11.56\pm0.06$ &  $3.7\pm0.9$ \\
\FeII\ $\lambda\,2600$ & $51\pm2$ & $13.11\pm0.05$ &$1.74\pm0.04$ \\
\FeII\ $\lambda\,2586$ & $29\pm2$ & $13.11\pm0.05$ & $1.74\pm0.04$ \\
\FeII\ $\lambda\,2382$ & $50\pm2$ & $13.11\pm0.05$ & $1.74\pm0.04$ \\
\FeII\ $\lambda\,2374$ & $14\pm2$ & $13.11\pm0.05$ & $1.74\pm0.04$ \\
\FeII\ $\lambda\,2344$ & $32\pm2$ & $13.11\pm0.05$ & $1.74\pm0.04$ \\
\FeI\ \ $\lambda\,3861$  & $4\pm1$ & $12.25\pm0.03$ & $0.8\pm0.1$ \\
\FeI\ \ $\lambda\,3720$ & $6\pm1$ &  $12.25\pm0.03$ & $0.8\pm0.1$ \\
\FeI\ \ $\lambda\,3441$ & $3\pm1$ &  $12.25\pm0.03$ & $0.8\pm0.1$ \\
\FeI\ \ $\lambda\,3021$ & $8\pm1$ &  $12.25\pm0.03$ & $0.8\pm0.1$ \\
\FeI\ \ $\lambda\,2984$ & $3\pm1$ &  $12.25\pm0.03$ & $0.8\pm0.1$ \\
\FeI\ \ $\lambda\,2967$ & $4\pm1$ &  $12.25\pm0.03$ & $0.8\pm0.1$ \\
\mbox{Li\,{\sc i}}\ \ \ $\lambda\,6709$ & $< 11$ &  & \\                  
\mbox{Ti\,{\sc ii}}\ \ $\lambda\, 3384$ & $<2$ & & \\  
\mbox{CH}\ \ \ $\lambda\,4301$ & $<3$ & & \\
\mbox{CH$^+$}\ \ $\lambda\,4223$ & $<3$ & & \\
\mbox{CN}\ \ \ $\lambda\,3875$ & $<3$ & & \\
\hline                  
\end{tabular}
\end{table}
%

In Table~\ref{linepar} we report for all the measured transitions the
redshift, the rest EW, the column densities, and Doppler parameters 
resulting from the Voigt fit with $1 \sigma$ uncertainties given by
Lyman in MIDAS.  
We give limits on the rest EWs of the other ionic and molecular
transitions that are generally observed 
in the local interstellar clouds and that fall in the observed
spectral range: 
\mbox{Ti\,{\sc ii}} $\lambda\, 3384$, \mbox{Li\,{\sc i}} $\lambda\, 
6709$, CH $\lambda\, 4301$, CH$^+$ $\lambda\, 4223$, and CN $\lambda\,
3875$ \AA. 


\subsection{Comparison with weak \MgII\ systems}


High-resolution QSO spectra have allowed the detection of a population
of weak \MgII\ systems \citep[with EWs $0.02 < 
W_o^{\lambda 2796}<0.3$ \AA, ][]{churchill99}. They are thought to arise in 
sub-Lyman limit systems ($15.8 < \log N($\HI$) < 16.8$), comprising 
at least 25 \% of \Lya\ forest clouds in that column density range
\citep{rigby02}.  
Weak \MgII\ absorbers are also found to have high metallicity, at
least 10 \% solar, but in some cases even solar or supersolar
\citep{rigby02,charlton03}. 

We have searched the 18 QSO spectra of  the ESO Large Programme 
(LP) for weak \MgII\ absorptions  in the redshift range $0.4 < z
<1.4$, excluding those falling in the \Lya\ forest. Adopting a nearest
neighbour velocity separation $> 500$ km s$^{-1}$  \citep[following
][]{churchill99}, we found 28 systems in a redshift path $\Delta
z=13.6$ \citep[see also ][]{lynch06}. 
We also examined the sample of \citet{churchill99} observed with HIRES
at Keck, in the same redshift range, at a similar
resolution, but in general at a lower signal-to-noise ratio. 

To compare the properties of our peculiar absorber with the
sample of weak \MgII\ systems, we measured the EWs
of \MgII\ $\lambda\, 2796$, \FeII\ $\lambda\,2600$, and \MgI\
$\lambda\,2852$ (when detected) for all velocity components at
separations $\ge 50$ km s$^{-1}$ (a total of 34 components) and
considered only single-component systems of the sample of  
\citet{churchill99} (24 systems).


   \begin{figure}
      \includegraphics[width=9truecm]{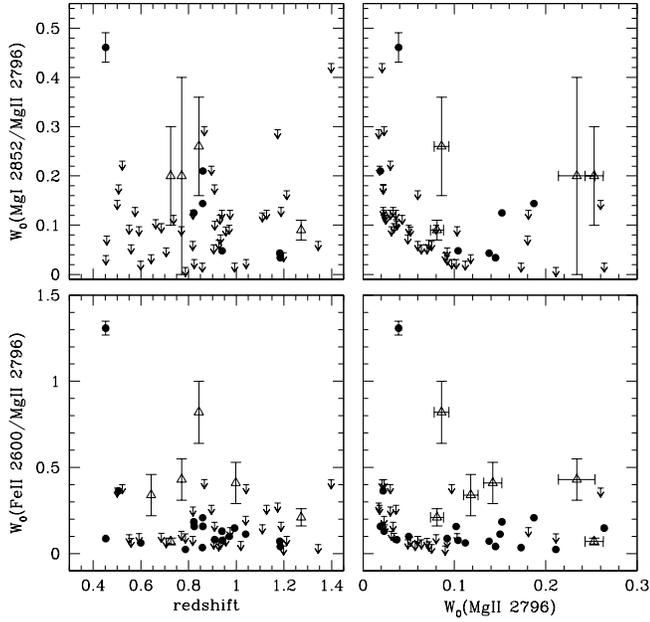}
     \caption{EW ratios \MgI($\lambda 2851$)/\MgII($\lambda 2796$) 
     and \FeII($\lambda 2600$)/\MgII($\lambda 2796$) of the single components for 
     the LP (dots) and \citet{churchill99} (open triangles) weak \MgII\ samples 
     as a function of redshift and of $W_o(\lambda 2796)$. The error
     bars on the dot in the upper left corner (corresponding to the
     systems that we are discussing)  
     are representative of the $1 \sigma$ uncertainties on our data. }
         \label{fig:weakmg2}
   \end{figure}
    
In Fig.~\ref{fig:weakmg2}, we plot the EW ratios 
\MgI($\lambda 2851$)/\MgII($\lambda 2796$) and \FeII($\lambda
2600$)/\MgII($\lambda 2796$) of the single components as a function 
of redshift and of $W_o^{\lambda 2796}$. 
The peculiar component of our system shows the highest value of the 
samples  for both EW ratios, significantly discrepant from 
the general distribution. 

This result and the presence of rare neutral species suggests that
this absorber does not reflect the general properties of the class of
weak \MgII\ system, but could instead trace a large amount of cold 
neutral hydrogen. 

\subsection{Comparison with DLA systems}

   \begin{figure}
      \includegraphics[width=9truecm]{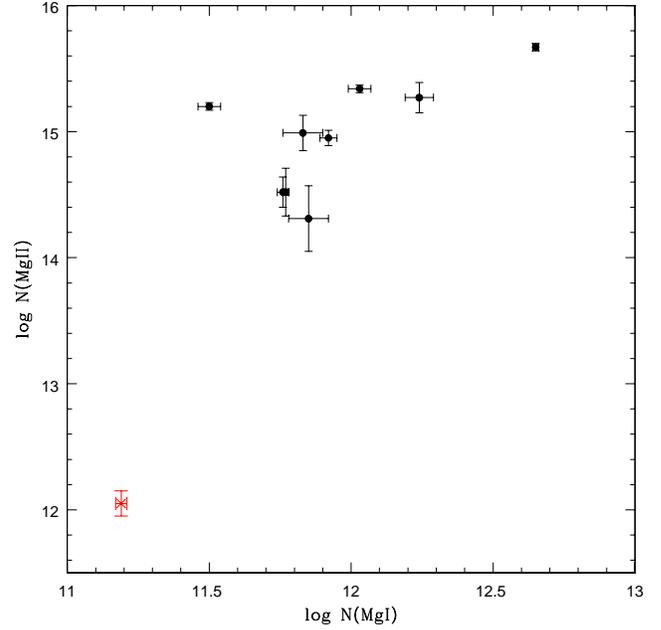}
     \caption{$\log N($\MgII$)$ vs. $\log N($\MgI$)$ for the sample of
     DLA components collected by \citet{mirka06} and for the studied
     system component (cross).}
         \label{fig:mg2vsmg1}
   \end{figure}

   \begin{figure}
      \includegraphics[width=9truecm]{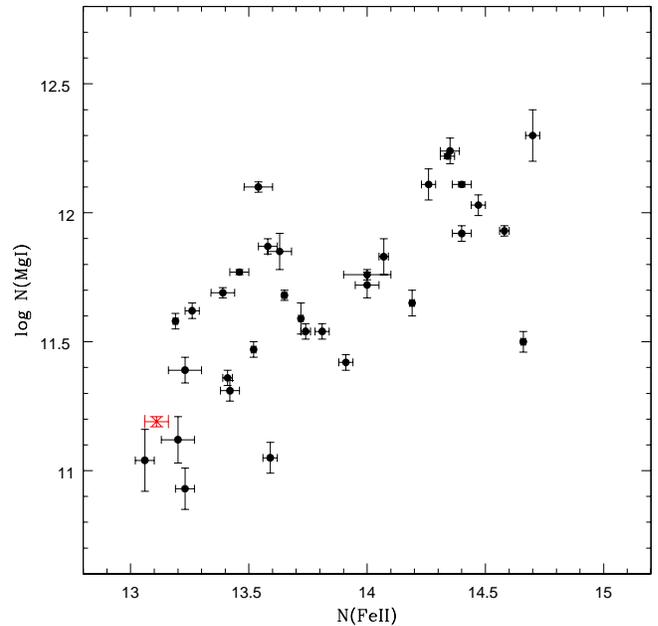}
     \caption{$\log N($\MgI$)$ vs. $\log N($\FeII$)$ for the sample of
    DLA components collected by \citet{mirka06} and for the studied
     system component (cross). }
         \label{fig:mg1vsfe2}
   \end{figure}

DLA systems are characterized by the highest neutral hydrogen column
density among QSO absorbers ($N($\HI$)\ge 2 \times 10^{20}$ \cm);
however, up to now there are no detections of \FeI, \CaI, and \SiI\ in
those systems.  Also, molecular hydrogen is not very common: it is
present in about 13-20 percent of the presently observed DLA systems 
\citep{srianand05}. The coldest environments, rich in molecules and
possibly in metals, could be missed by observations due to dust obscuring 
the background source.   

We have compared the properties of our absorption system with the
sample of 11 DLA by \citet{mirka06}, where ionic column 
densities were measured component by component. Considering those
components for which the column density were measured for at
least two ions among \MgI, \MgII, \MnII, and \FeII, we built a sample
of 58 components. 
Figure~\ref{fig:mg2vsmg1} shows that, while the amount of \MgI\
observed in our system is comparable with that of the examined
sample of DLA (see also Fig.~\ref{fig:mg1vsfe2}), there is a
difference of more than 2 orders of magnitude in the \MgII\ column
density, which, as shown in Sect.~3.1 cannot be ascribed to
measurement uncertainties.  On the other hand, we tend to exclude
the possibility that \mbox{Mg\,{\sc iii}} is the dominant ionization
state in this system, due to the presence of the neutral elements. 
Now, if we look at Fig.~\ref{fig:mg1vsfe2}, we see that the measured
\FeII\ column density is compatible with the lower values measured in
the considered sample of DLA components.  
Note that in this plot there is a larger number of \MgI\ measurements
than in Fig.~\ref{fig:mg2vsmg1}, which is due to the fact that many
systems with measured \MgI\ and \FeII\ column densities have
saturated (thus not measured) \MgII\ lines.  

The observed difference in the \MgI/\MgII\ ratios could be
due to a different ionization state in our system than the DLA
population. However, the difficulty in 
measuring a precise column density for single \MgII\ components in DLA
due to saturation or blending may also play an important role.  

Supposing that no ionization corrections are needed for our system,
the abundance ratio with respect to solar of magnesium and iron
(taking into account the contribution of all ionization stages) is
[Mg/Fe]~$\simeq -1.05\pm0.05$, while the mean value for the considered
sample of 11 DLA is $0.7\pm 0.4$.  On the other hand, [Mn/Fe]~$\simeq 
0.31\pm0.05$ at a variance with the DLA mean value $\simeq -0.23 \pm
0.05$.   


\subsection{Comparison with local interstellar clouds} 

Another class of absorbers characterized by high neutral hydrogen
column densities and showing some of the neutral transitions that we
have measured is represented by the local cold interstellar clouds
\citep[][ WHM, and references therein]{welty03}.  


We compared the properties of our system at $z\simeq 0.4521$ with 
the  column densities for \CaI, \CaII, and \FeI\ of local lines of sight towards 
star forming regions in Table~9 of WHM (excluding upper limits and uncertain 
determinations), while the \HI\ column densities for the corresponding
lines of sight are taken from \citet[][ and references
  therein]{welty_hobbs01}.       

   \begin{figure}
      \includegraphics[width=9truecm]{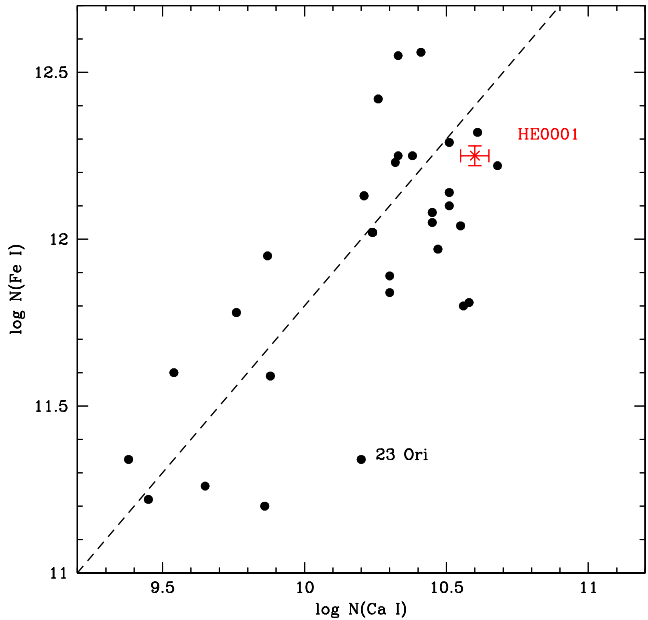}
     \caption{$\log N($\FeI$)$ vs. $\log N($\CaI$)$ for the sample of
     interstellar diffuse clouds collected in \citet{welty03} and for
     our QSO absorption system (cross). The dashed line
     indicates a linear relationship between the two species.}
         \label{fig:fe1vsca1}
   \end{figure}

In Fig.~\ref{fig:fe1vsca1} the column densities of \FeI\
vs. \CaI\ measured for the local interstellar medium are plotted with
those of our absorber.  
The high-redshift value is in very good agreement with the
local measurement and resides at the high column density tip of the
distribution. 
The local absorption systems have neutral hydrogen column densities 
varying approximately between $\log N($\HI$) \sim 20.5$ and $21.5$,
with a slight correlation with the \CaI\ and \FeI\ column densities,
indicating a high \HI\ column  density for our absorber. 
Allowing for that range of \HI\ column densities and taking the observed
total column density of (neutral plus singly ionized) Fe in our
system, we estimated a range of metallicities, $-2.78 \la$~[Fe/H]~$\la
-3.78$, lower than the lowest metallicities observed in DLA. At these
Fe abundances, the effect of depletion onto dust grains should
be negligible.

   \begin{figure}
      \includegraphics[width=9truecm]{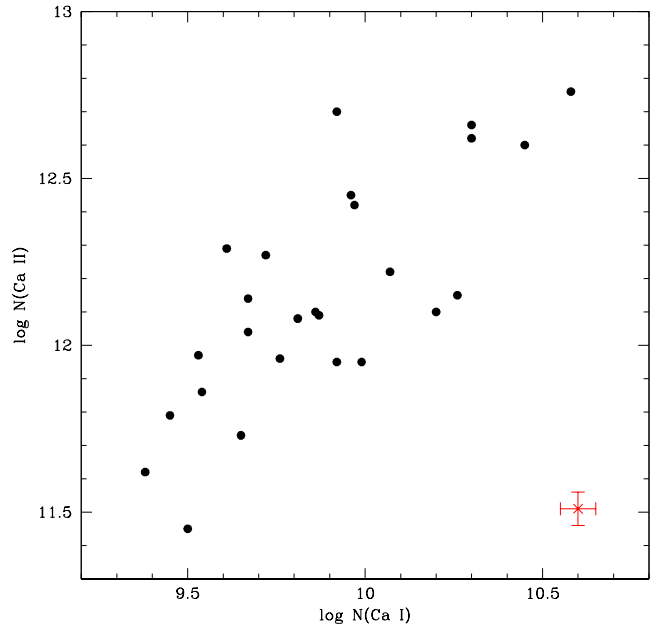}
     \caption{$\log N($\CaII$)$ vs. $\log N($\CaI$)$ for the sample of
     interstellar diffuse clouds collected in \citet{welty03} and for
     the studied QSO absorption system (cross).}
         \label{fig:ca2vsca1}
   \end{figure}

Figure~\ref{fig:ca2vsca1} shows $\log N($\CaII$)$ vs. $\log N($\CaI$)$
for the local sample and for our system.  As in the case of
\MgII\ absorptions in DLA (see Fig.~\ref{fig:mg2vsmg1}), there is a
deficiency of \CaII\ in our system with respect to the general
distribution of interstellar clouds. 
A detailed analysis of single-component \CaI\ and \CaII\ column
densities in stellar spectra at very high resolution ($FWHM\sim
0.6-1.8$ km s$^{-1}$), shows that the average single component column
density ratio is \CaI/\CaII~$\la -2$ \citep{pan04}. This result is a
strong indication that 
the difference in \CaII\ abundance between our system and the local
interstellar clouds is due to a different ionization state.

\section{Photoionization model}

From the comparisons carried out in the previous sections, it is
clear that our system shows some characteristics close to those of
high \HI\ column density absorbers. On the other hand, its ionization
properties look different both from those of DLA and of local
interstellar clouds.  

We ran photoionization models with the version c06.02c of Cloudy
\citep{ferland} to reproduce the observed ionization properties and
investigate the physics of the studied cloud. 
We considered two types of radiation fields: a hard, QSO dominated
spectrum representative of the UV background external to the 
system \citep[as modeled by][ at $z=0.49$]{hm99}, and a soft, 
stellar-type spectrum \citep[$T_{\rm eff}=33,000$ K, $\log
  g=4$;][]{kurucz} representative of the internal radiation field or of 
an external field dominated by starlight from galaxies.
A solar abundance pattern\footnote{Solar abundances in 
Cloudy are taken from \citet{holweger01} for Mg and Fe and
from \citet{GS98} for Ca} has been assumed and a metallicity,
$Z$, computed from the observed Fe total column density and the
adopted \HI\ column density. 
The total density, thickness of the gas slab and \HI\ column density
are strictly related, $n_{\rm H} \times \Delta R \simeq
N($\HI$)$, since the investigated cloud is dominated by neutral gas. The
first two parameters are constrained by observations.  
An upper limit on the size of the cloud of $\sim
150-200\,h^{-1}_{71}$ pc, comes from the study of single \MgII\
components in spectra of lensed QSOs \citep{rauch02}. 
Molecular clouds associated with DLA in QSOs spectra have total
densities $n_{\rm H} \simeq 10-200$ cm$^{-3}$ and temperatures
$T\simeq 100-300$ K \citep{srianand05}, while, local cold interstellar
clouds can be as small as a few parsecs, with total densities $n_{\rm
  H} \sim 10-15$ cm$^{-3}$ and temperature $T\sim 100$ K
\citep[e.g.][]{welty99}.   

Taking the previous constraints into account, we ran a grid of photoionization
models varying  $n_{\rm H}$, $N($\HI$)$, $Z$ and the ionization
parameter

\begin{equation}
U \equiv \frac{\Phi_{912}}{c\, n_{\rm H}} = \frac{4\, \pi\,
  J_{912}}{h\,c\,n_{\rm H}} \simeq 2 \times 10^{-5}\,
  \frac{J_{912}/10^{-21.5}}{n_{\rm H}/1\, \mbox{cm}^{-3}}, 
\end{equation}

\noindent
where $J_{912}$ is the intensity of the ionizing spectrum at the Lyman
limit.

\begin{table*}
\begin{minipage}[t]{\textwidth}
\renewcommand{\footnoterule}{}
\caption{Characteristic parameters and results of the photoionization
  models (see text).}
\label{photomodel}      
\centering 
\begin{tabular}{l l c c r r c c c c c c}
\hline\hline
Model & Spectrum & $\log N($\HI$)$ & $n_{\rm H}$ & $\Delta R$ & $Z$ &
$\log U_{\rm Mg}$ & $\log J_{\rm Mg}$ & $\log U_{\rm Ca}$ & 
$\log J_{\rm Ca}$ & (\FeI/\FeII)$_{\rm max}$ & T$_{\rm min}$\\
 & & cm$^{-2}$ & cm$^{-3}$ & pc & & ($\pm 0.03$) & ($\pm 0.03$)& ($\pm
0.05$)& ($\pm 0.05$)&& K \\
\hline \\
HM1 & hard & 20.7 & 10.0  & 15.0 & -3.0 & -6.62 & -22.42 & -7.56 &
-23.36 & -2.81 & 110 \\
STAR1 & soft & 20.7 & 10.0  & 15.0 & -3.0 & -6.66 & -22.46 & -7.36 &
-23.16 & -2.80 & 110 \\
HM2 & hard & 21.5 & 10.0  & 102.5 & -3.8 & -6.66 & -22.46 & -7.76 &
-23.56 & -2.81 & 111 \\
STAR2 & soft & 21.5 & 10.0  & 102.5 & -3.8 & -6.65 & -22.45 & -7.54 &
-23.34 & -2.81 & 111 \\
HM3 & hard & 21.5 & 63.1  & 15.0 & -3.8 & -6.95 & -21.95 & -7.93 &
-22.93 & -2.71 & 83 \\
STAR3 & soft & 21.5 & 63.1  & 15.0 & -3.8 & -6.95 & -21.95 & -7.71 &
-22.71 & -2.71 & 83 \\
HM4 & hard & 20.7 & 158.5 & 1.0 & -3.0 & -7.08 & -21.68 & -8.01 &
-22.61 & -2.64 & 67 \\
STAR4 & soft & 20.7 & 158.5 & 1.0 & -3.0 & -7.11 & -21.61 & -7.80 &
-22.40 & -2.64 & 66 \\
HM5 & hard & 21.5 & 1000.0 & 1.0 & -3.8& -7.47 & -21.27 & -8.40 &
-22.20 & -2.61 & 62 \\ 
STAR5 & soft & 21.5 & 1000.0 & 1.0 & -3.8 & -7.49 & -21.29 & -8.19 &
-21.99 & -2.62 & 61 \\
\hline
\end{tabular}
\footnotetext{The observed values for the column
  density ratios are: \MgI/\MgII~$\simeq -0.96\pm0.03$,
  \CaI/\CaII~$\simeq -0.91\pm0.05$ and \FeI/\FeII~$\simeq
  -0.86\pm0.04$.}
\end{minipage}
\end{table*}



\begin{figure}
      \includegraphics[width=9truecm]{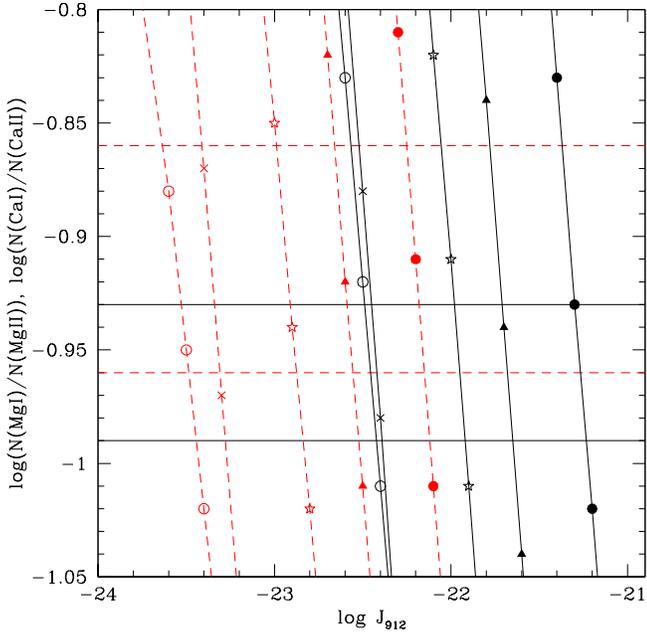}
     \caption{$\log (N($\MgI$)/N($\MgII$))$ (solid lines) and $\log
     (N($\CaI$)/N($\CaII$))$ (dashed lines) as a function of the log of the
     Lyman limit intensity of the ionization spectrum for the 5 test
     photoionization model run with Cloudy (see
     Table~\ref{photomodel}) in the case of the hard spectrum. The
     legend of symbols is the following: cross - Model 1, empty circle
     - Model 2, star - Model 3, solid triangle - Model 4, solid circle
     - Model 5. The solid and dashed horizontal lines delimit  
     the 1~$\sigma$ interval of the observed values for $\log
     (N($\MgI$)/N($\MgII$))$ and $\log (N($\CaI$)/N($\CaII$))$, 
     respectively.} 
         \label{fig:photoioniz}
   \end{figure}

\begin{figure}
      \includegraphics[width=9truecm]{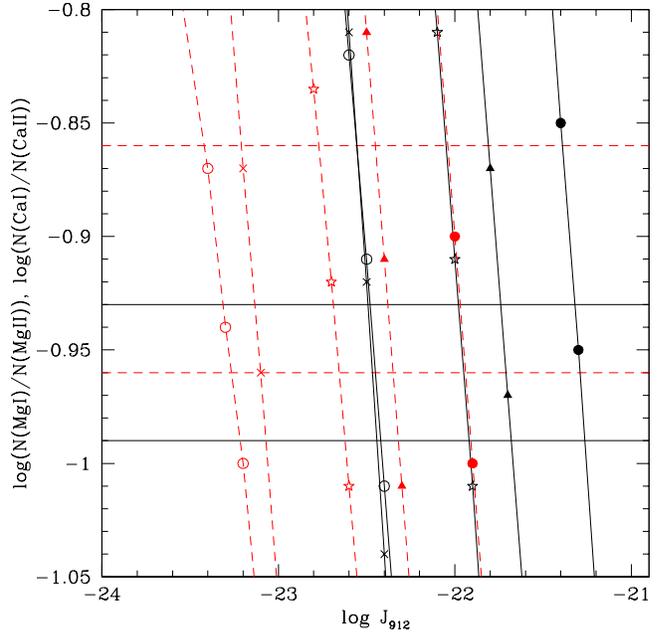}
      \caption{Same as Fig.~\ref{fig:photoioniz} but for the stellar
      ionizing spectrum.}
      \label{fig2:photoioniz}
\end{figure}

The parameters of the five physical models that we studied
  in detail, adopting both the hard and soft ionizing spectra, are
reported in Table~\ref{photomodel}, together with the values of the
ionization parameter, $U$, and the corresponding ionizing flux at the 
Lyman Limit, $J$, at which we obtained the observed values for $\log
     (N($\MgI$)/N($\MgII$))$ and $\log (N($\CaI$)/N($\CaII$))$. In
columns 11 and 12, we also report the maximum value obtained for the
$\log (N($\FeI$)/N($\FeII$))$ ratio and the corresponding minimum
temperature reached for that model. We stopped when a further decrease
in $U$ did not correspond to a decrease in $T$. 

Model by model, the main difference due to the different ionizing
spectrum is the value of the ionization parameter at which we obtained 
the observed column density ratio for Ca, which is a factor $\sim 1.6$ 
systematically higher for the soft spectrum.
As expected, the lower temperatures are reached in the region
far away from the ionizing source and strongly depend on the
adopted total density. The Fe column density ratio is tightly 
related to the lowest temperature reached (about 60 K), and we could
not obtain values of $\FeI/\FeII$ higher than $\simeq 0.0024$ or about 55
times lower than the observed value.
In Figs.~\ref{fig:photoioniz}, \ref{fig2:photoioniz} we report the obtained $\log
N($\MgI$)/N($\MgII$)$ and $\log N($\CaI$)/N($\CaII$)$ as a function
of the intensity at the Lyman limit of the hard and soft ionizing
spectra. 
Higher intensities are needed for denser clouds. There is a range of
intensities for both ionizing spectra, $-22.45\la \log J_{912}\la
-21.95$, at which the observed $\log N($\MgI$)/N($\MgII$)$ and 
$\log N($\CaI$)/N($\CaII$)$ are reproduced. 
Neutral and singly ionized Mg arise in clouds with total
density $n_{\rm H}=10$ or $\sim 63.1$ cm$^{-3}$ and Ca in clouds
with $n_{\rm H}\simeq 158.5$ or $1000$ cm$^{-3}$, respectively. 
The above range of $J_{912}$ agrees closely with the intensity of the
UV background measured at $z<1$ from the QSO proximity effect, $-22.3
\la \log J_{912}\la -21.98$ \citep{scott}.

For comparison, the average $\log N($\CaI$)/N($\CaII$)$ ratio observed
for local interstellar clouds is obtained for $\log J_{912}\sim -22$
in the case of a cloud of 15 pc and $n_{\rm H}=10$ cm$^{-3}$. It is
interesting to note that the 2nd main ionization state of Ca in
this conditions is \mbox{Ca\,{\sc iii}}, while at the ratio observed
in our system it is \CaI. 

The main result of our calculations is that we could not find any
combination of the parameters resulting in a model that could
reproduce the three observed ratios at the same time:
$N($\MgI$)/N($\MgII$)$, $N($\CaI$)/N($\CaII$)$, and
$N($\FeI$)/N($\FeII$)$.  We can infer that the UV background is 
likely to be the ionizing source of the observed gas; however, simple 
photoionization models with uniform density cannot explain the
observations and more complex density structures 
should be adopted. 


\section{Conclusions}

We report the study of a very peculiar metal absorption system
detected in the UVES spectrum of the QSO HE0001-2340 at $z\simeq 0.452$. 

The system is characterized by a single narrow component in \MgII\ and
is classified as a weak \MgII\ absorber. However, at the same redshift
we also observed transitions due to \MgI, \CaII, \MnII, \FeII\ and, in
particular, to \SiI, \CaI, \FeI, which are observed for the  
first time in a QSO absorption system. 
The difference between the properties of our system and those of the
sample of weak \MgII\ systems that we have collected  in the ESO Large
Programme QSO spectra gives an estimate of the probability of 
intersecting such a cloud: we observed one over 34 (single \MgII\
components), or $P\approx 0.03$.    

The presence of rare neutral elements  suggests that the gas is
shielded by a large amount  
of neutral hydrogen, so we have compared its chemical properties with
those of the highest \HI\ column density QSO absorptions, the damped
\Lya\ systems,  and with the local cold interstellar clouds.  
The main results deriving from this comparison are the following. 
\par\noindent
1. The ratios \MgI/\MgII\  and \CaI/\CaII\  are at least two and
  one orders of  magnitude higher in our system than in the other absorbers. 
A large fraction of doubly ionized Mg and Ca is excluded
by the presence of the rare neutral species. It has to be noted that
there are only a few measures of \MgII\ in DLA \citep[see][]{mirka06};
and in general, both in DLA and local interstellar clouds, \MgII\ and
\CaII\ are more extended than the corresponding \MgI\ and \CaI, implying that
they are due to more extended regions. 
However, in very high-resolution stellar spectra, the \CaI/\CaII\  ratio computed 
component-by-component is still one order of magnitude lower than in our system, 
strongly suggesting that our system has a lower ionization state. 
\par\noindent
2. We can give a rough estimate of the metallicity by assuming that our
system has an amount of \HI\ comparable to what observed
in local interstellar clouds with similar abundances of \CaI\ and
\FeI. This implies a range  of metallicities $-2.78 \la$~[Fe/H]~$\la -3.78$, lower
than the lowest metallicities measured in DLA.    
\par\noindent
3. More puzzling is the underabundance of Mg with respect to
Fe, which is very hard to explain both with nucleosyntesis and with
differential dust depletion. Indeed, interstellar cold clouds 
generally show significantly larger Fe depletion than Mg depletion
\citep[e.g.][]{savage96}.  

We studied the physical and ionization properties of our cloud
by running a grid of photoionization models with Cloudy. It is not
possible to recover the observed \MgI/\MgII, \CaI/\CaII, and
\FeI/\FeII\ column density ratios with a single gas slab of constant
density. On the other hand, by adopting an ionizing spectrum compatible with
the UV background at $z<1$, the correct Mg and Ca ratios
are obtained in gas with a total density of $n_{\rm H}\sim10-60$ and $150-1000$
cm$^{-3}$, respectively.  
The Fe ratio cannot be reproduced, but there are indications that 
gas denser than $n_{\rm H}\sim 1000$ cm$^{-3}$ is needed. 
These results suggest that only a complex density structure for the cloud 
could explain the observed ionic abundances.

What is this very rare cloud? A definite answer could come from UV
observations of this object to measure the \HI\ and the H$_2$.     
Furthermore, a better search for these class of systems would be
possible with spectrographs at a higher resolving power ($>100\,000$) than
presently provided by instruments like UVES and HIRES on 8-10m class
telescopes.  

\begin{acknowledgements}
It is a pleasure to thank M. C\'enturion, S. Cristiani,
S. D'Odorico, P. Molaro, and G. Vladilo for enlightening
discussions. 
We are grateful to the referee, whose comments and suggestions greatly
improved the quality of the paper.  
\end{acknowledgements}

\bibliographystyle{aa}
\bibliography{aamnem99,myref}

\end{document}